\documentclass[%
 reprint,
superscriptaddress,
longbibliography,
 amsmath,amssymb,
pre,
]{revtex4-2}

\usepackage{lipsum}
\usepackage{graphicx}% Include figure files
\usepackage{dcolumn}% Align table columns on decimal point
\usepackage{bm}% bold math
\usepackage{hyperref}% add hypertext capabilities
\usepackage[mathlines]{lineno}% Enable numbering of text and display math
\newcommand*\patchAmsMathEnvironmentForLineno[1]{%
  \expandafter\let\csname old#1\expandafter\endcsname\csname #1\endcsname
  \expandafter\let\csname oldend#1\expandafter\endcsname\csname end#1\endcsname
  \renewenvironment{#1}%
     {\linenomath\csname old#1\endcsname}%
     {\csname oldend#1\endcsname\endlinenomath}}% 
\newcommand*\patchBothAmsMathEnvironmentsForLineno[1]{%
  \patchAmsMathEnvironmentForLineno{#1}%
  \patchAmsMathEnvironmentForLineno{#1*}}%
\AtBeginDocument{%
\patchBothAmsMathEnvironmentsForLineno{equation}%
\patchBothAmsMathEnvironmentsForLineno{align}%
\patchBothAmsMathEnvironmentsForLineno{flalign}%
\patchBothAmsMathEnvironmentsForLineno{alignat}%
\patchBothAmsMathEnvironmentsForLineno{gather}%
\patchBothAmsMathEnvironmentsForLineno{multline}%
}
\newcommand{\expct}[1]{\langle #1 \rangle}
\newcommand{\figref}[1]{Fig.\,\ref{#1}}
\newcommand{\figsref}[1]{Figs.\,\ref{#1}}
\renewcommand{\eqref}[1]{Eq.\,(\ref{#1})}
\newcommand{\tbref}[1]{Table\,\ref{#1}}

\newcommand{\vidref}[1]{Movie~#1}
\newcommand{\vidsref}[1]{Movies~#1}
\newcommand{\supfigref}[1]{Fig.\,\ref{#1}}

\newcommand{\unit}[1]{\,\mathrm{#1}}
\newcommand{\degc}{\unit{{}^\circ{}C}}

\usepackage{xr} %external reference
\makeatletter
\newcommand*{\addFileDependency}[1]{
  \typeout{(#1)}
  \@addtofilelist{#1}
  \IfFileExists{#1}{}{\typeout{No file #1.}}
}
\makeatother

\newcommand*{\myexternaldocument}[1]{
    \externaldocument[S-]{build/#1}  % for arXiv
    \addFileDependency{#1.tex}
    \addFileDependency{build/#1.aux}  % for arXiv
}
\myexternaldocument{BDpaperSI-arxiv}  % for arXiv

\begin{document}

%\title{bundle formation in planktonic \textit{E. coli} populations under nutrient starvation}
\title{Smectic-like bundle formation of planktonic bacteria upon nutrient starvation}

\author{Takuro Shimaya}
\email{t.shimaya@noneq.phys.s.u-tokyo.ac.jp}
\affiliation{Department of Physics,\! The University of Tokyo,\! 7-3-1 Hongo,\! Bunkyo-ku,\! Tokyo 113-0033,\! Japan}%
\author{Kazumasa A. Takeuchi}
\email{kat@kaztake.org}
\affiliation{Department of Physics,\! The University of Tokyo,\! 7-3-1 Hongo,\! Bunkyo-ku,\! Tokyo 113-0033,\! Japan}%
\affiliation{Institute for Physics of Intelligence,\! The University of Tokyo,\! 7-3-1 Hongo,\! Bunkyo-ku,\! Tokyo 113-0033,\! Japan}%

\date{\today}

\begin{abstract}
Bacteria aggregate through various intercellular interactions to build biofilms, but the effect of environmental changes thereupon remains largely unexplored.
Here, by using an experimental device that overcomes past difficulties, we observed the collective response of \textit{Escherichia coli} aggregates against dynamic changes in the growth condition.
We discovered that nutrient starvation caused bacterial cells to arrange themselves into bundle-shaped clusters, developing a structure akin to that of smectic liquid crystal.
The degree of the smectic-like bundle order was evaluated by a deep learning approach.
Our experiments suggest that both the depletion attraction by extracellular polymeric substances and the growth arrest are essential for the bundle formation.
Since these effects of nutrient starvation at the single-cell level are common to many bacterial species, bundle formation might also be common collective behavior that bacterial cells may exhibit under harsh environments. 
\end{abstract}
\maketitle

\section{Introduction}
Under favorable environments, bacteria may proliferate by self-replications and eventually build densely organized communities, in particular biofilms \cite{Flemming2016}.
Biofilms are widely distributed in various habitats, e.g., ocean, soil, kitchens, and also we humans grow bacteria in our intestines to live in harmony with each other \cite{devos2015}.
Biofilms are also involved in our daily lives and industries, helping to degrade organic waste and produce fine chemicals \cite{Halan2012}.
While biofilms benefit humanity in this way, they also cause a variety of problems in, e.g., medical equipment \cite{Shirtliff2009} and industrial products \cite{MattilaSandholm1992}.
Therefore, understanding the physiology of biofilms is a crucial mission across diverse disciplines.

During the typical biofilm formation process, bacteria aggregate through a variety of cell-cell interactions \cite{Trunk2018}.
While cell adhesion to material surfaces is regarded as one of the most crucial events in biofilm formation \cite{Romeo2008,Palmer2007}, cell-cell interactions mainly promote aggregation of planktonic cells that are not attached to the surface, which has also been shown to be important for biofilm construction \cite{Kragh2016}.
In the case of \textit{Escherichia coli}, the Ag43 antigen protein plays a major role in the aggregation process \cite{Hasman1999,Kjaergaard2000,Danese2000,Trunk2018} via protein-protein interactions \cite{Heras2014}.
%, which is known to be maintained by protein-protein interactions between Ag43 molecules \cite{Heras2014}.
Extracellular polysaccharides and other debris also contribute through depletion attraction \cite{Dorken2012,Amimanan2017,Secor2018,Porter2019,Kanlaya2019,Secor2021}.
Type I pili are also known to be important in some cases \cite{Pratt1998,Danese2000,Schembri2001}.
%It is also known that aggregation is mediated by the depletion attraction of macromolecules such as extracellular polysaccharides and other debris \cite{Dorken2012,Amimanan2017,Secor2018,Porter2019,Kanlaya2019,Secor2021}.

The expression state of extracellular molecules involved in cell-cell interactions, including Ag43, is generally highly dependent on the environment.
Interestingly, it was reported that the relative importance of Ag43 and type I pili in aggregation depends on the environment: Ag43 has less effect on aggregation in Luria–Bertani (LB) broth than in minimal medium such as M9 \cite{Danese2000}.
%For example, it has been suggested that both Ag43 and type I pili contribute to aggregation are mutually exclusive depending on the environment; an increase in the expression of one of them inhibits the effect of the other on aggregation \cite{Hasman1999,Diderichsen1980}.
%For example, when bacterial cells are cultured in LB broth, Ag43 has less effect on aggregation than in minimal medium such as M9 \cite{Danese2000}, and it has also been argued that type I pili itself may also be crucial for aggregation processes \cite{Pratt1998,Schembri2001}.
Besides, the secretion of proteins and polysaccharides, known as extracellular polymeric substances (EPS), is known to be sensitive to environmental changes such as nutrient starvation \cite{Myszka2009,Myszka2011,Zhang2014,Costa2018}.

Because of the sensitivity of cell-cell interactions to the environment, aggregates of planktonic cells may collectively respond to environmental changes.
However, planktonic aggregates under time-dependent environments have been hardly investigated.
A major difficulty is the necessity to change the environment in a controlled manner, maintaining spatial uniformity without adding hydrodynamic perturbations, which may influence the aggregates \cite{Karimi2015,Wu2016,Pousti2019} and even destroy in some cases.
%This requirement is difficult to achieve in typical microfluidic devices \cite{Karimi2015,Wu2016,Pousti2019} made of polydimethylsiloxane (PDMS).

Here, we overcome the difficulty by a membrane-type microfluidic device that we developed previously, namely the extensive microperfusion system (EMPS), which can uniformly change the environment surrounding bacterial populations without hydrodynamic perturbations \cite{Shimaya2021}.
With this device and using several strains of \textit{E. coli}, we observed how planktonic aggregates respond against abrupt nutrient starvation. 
We discovered that, upon starvation, nematically ordered planktonic cells arranged themselves into bundle-shaped clusters, developing a structure akin to that of smectic liquid crystal. 
By employing a machine learning technique, we quantified the degree of this smectic-like bundle order. 
Moreover, we found that these bundles were destroyed if cell growth was resumed by switching back to the nutritious condition.
We also found that inhibition of EPS secretion hampered the bundle formation significantly. 
These results suggest that both depletion attraction between cells, mediated by secreted EPS, and the arrest of cell elongation are involved in the bundle formation. 

\begin{figure*}[t]
       \includegraphics[width=\hsize]{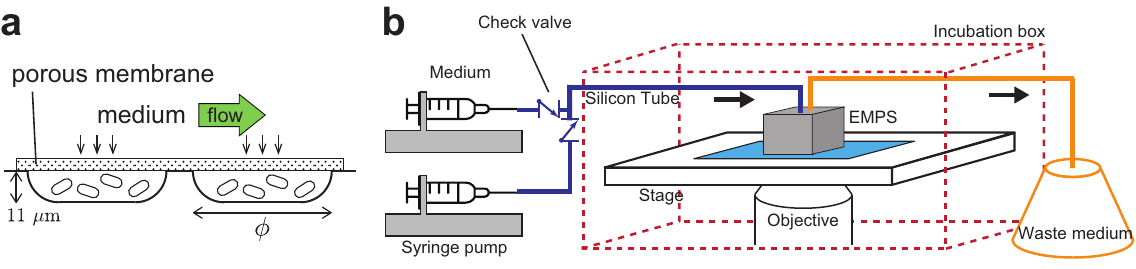}
       \caption{
       \label{fig-1}
       Experimental setup.
       (a) A cross-sectional illustration of EMPS.
       Bacterial cells were captured in wells (depth $11\unit{\mu{}m}$ and diameter $\phi$).
       The wells were covered by a rigid porous membrane, through which fresh medium was uniformly supplied.
       See Ref.~\cite{Shimaya2021}.
       (b) Overall setup of the experimental system.
       Liquid medium was constantly supplied from a syringe by a pump and drained into a waste container.
       By using another syringe with non-nutritious buffer, connected by a three-way junction with check valves, we can abruptly change the environment during observation.
       }
\end{figure*}

\section{Results}
\subsection{Experimental setup and observations under steady growth conditions}

Using EMPS (\figref{fig-1}), we observed planktonic aggregates of non-motile \textit{E. coli} (strain W3110 \cite{Jishage.Ishihama-JB1997}), formed in microfabricated wells on the substrate of the device.
The absence of motility is important to prevent motility-induced dispersion of cells.
The wells are circular and $11\unit{\mu{}m}$ deep, with diameters $\phi$ ranging from $110\unit{\mu{}m}$ to $330\unit{\mu{}m}$ (\figref{fig-1}a), fabricated on a coverslip by the method described in Ref.~\cite{Shimaya2021}.
Bacteria were confined in those wells by a porous membrane, attached to the substrate via the biotin-streptavidin bonding.
By supplying fresh medium through the porous membrane, we can control the culture condition in a spatially uniform manner, without hydrodynamic perturbations \cite{Shimaya2021}.
The device was placed in an incubation box maintained at $37\degc$, and fresh medium was automatically supplied by a syringe pump (\figref{fig-1}b).

We first observed the formation of planktonic aggregates under steady growth conditions (see Materials and Methods).
When the wild-type \textit{E. coli} W3110 (denoted by WT; see \tbref{tb-1} for the list of strains used in this work) was cultured in M9 minimum medium with glucose and amino acid [denoted by M9(glc.+a.a.); see \tbref{tb-2} for the list of measurements], we observed planktonic aggregates, as well as chains of cells that apparently divided but remained connected (\supfigref{S-figS1}a and \vidref1).
Similar chains were also observed in earlier studies \cite{Vejborg2009,Frommel2013,Puri.Allison-PNAS2024} and are known to be maintained by Ag43 localized on the pole of the cell \cite{Vejborg2009,Puri.Allison-PNAS2024}.
Using a strain W3110 \textit{flu}::\textit{kan} (hereafter called \textit{flu}::\textit{kan} mutant),
%(\textit{flu}::\textit{kan} mutant in \tbref{tb-1}),
for which the \textit{flu} gene producing Ag43 was replaced with a kanamycin-resistant gene (\textit{kan}), we confirmed that chain formation hardly appeared in this case and planktonic aggregates were mainly produced (\supfigref{S-figS1}b and \vidref2).
This indeed indicates the relevance of Ag43 to the chain formation.

We also observed that some cells adhered to the substrate in both the WT and \textit{flu}::\textit{kan} strains (see, e.g., \vidref2).
To focus on planktonic aggregations without adhesion of cells, here we attempted to use a strain without flagella, which are known to be an important factor for adhesion of \textit{E. coli} cells to surfaces \cite{Prigent2000,Friedlander2015,Kimkes2019}.
%While various molecules are involved in adhesion of \textit{E. coli} cells to surfaces \cite{Palmer2007}, it has been reported that the attachment of flagella to surfaces is one of the most important factors \cite{Prigent2000,Friedlander2015,Kimkes2019}.
Specifically, we used a mutant W3110 $\Delta$\textit{fliC} (hereafter called $\Delta$\textit{fliC} mutant), for which the \textit{fliC} gene producing flagella was deleted, and confirmed that adhesion of cells to the surface was significantly prevented (\supfigref{S-figS1}c) while planktonic aggregates appeared to be almost identical to those of the WT and the \textit{flu}::\textit{kan} mutant (\vidref3). 
We also found that chain formation was significantly suppressed in this case (\supfigref{S-figS1}c).
%Note that, to our knowledge, the relevance of flagella to chain formation has not been discussed.
These results imply that flagella and Ag43 jointly contribute to the chain formation, in agreement with a recent report on similar chain formation observed for another strain of \textit{E. coli} \cite{Puri.Allison-PNAS2024}.
%, though further validations are needed for a more decisive conclusion.
On the basis of these observations, in the following we will mainly use the $\Delta$\textit{fliC} mutant, to avoid both surface adhesion and chain formation.
%In any case, since we can avoid both surface adhesion and chain formation, in the following we will mainly use the $\Delta$\textit{fliC} mutant.
%, which allows us to intensively observe the planktonic aggregates.
With this strain, we measured the time evolution of the total area of aggregates in microscope images and verified that the growth rate of the aggregate size is comparable to the single-cell growth rate (\supfigref{S-figS1}d).
This indicates that the presence of aggregates does not deteriorate local growth conditions in the EMPS.  

\subsection{Bundle formation under nutrient starvation}

%As we already mentioned, various cell-cell interactions involved in aggregation processes are commonly dependent on the environment.
%For example, in the case of \textit{E. coli}, the secretion rate of cellulose released from cells into the environment \cite{Gualdi2008,Serra2013}, curli (amyloid fiber) growing on the cell surface \cite{Barnhart2006,Brombacher2006}, and lipopolysaccharide (LPS) constituting the extracellular membrane \cite{huisman1996} increases under external stresses through intracellular signals.
%In contrast, the secretion rate of some other substances, such as colanic acid \cite{Ionescu2009} and polyglucosamine, decreases under stresses \cite{Mika2014}.
%For \textit{P. aeruginosa}, it has directly been reported that nutrient starvation induces structural deformation of biofilms \cite{Hunt2004,Huynh2012}.

\begin{figure*}[p]
       \includegraphics[width=\hsize]{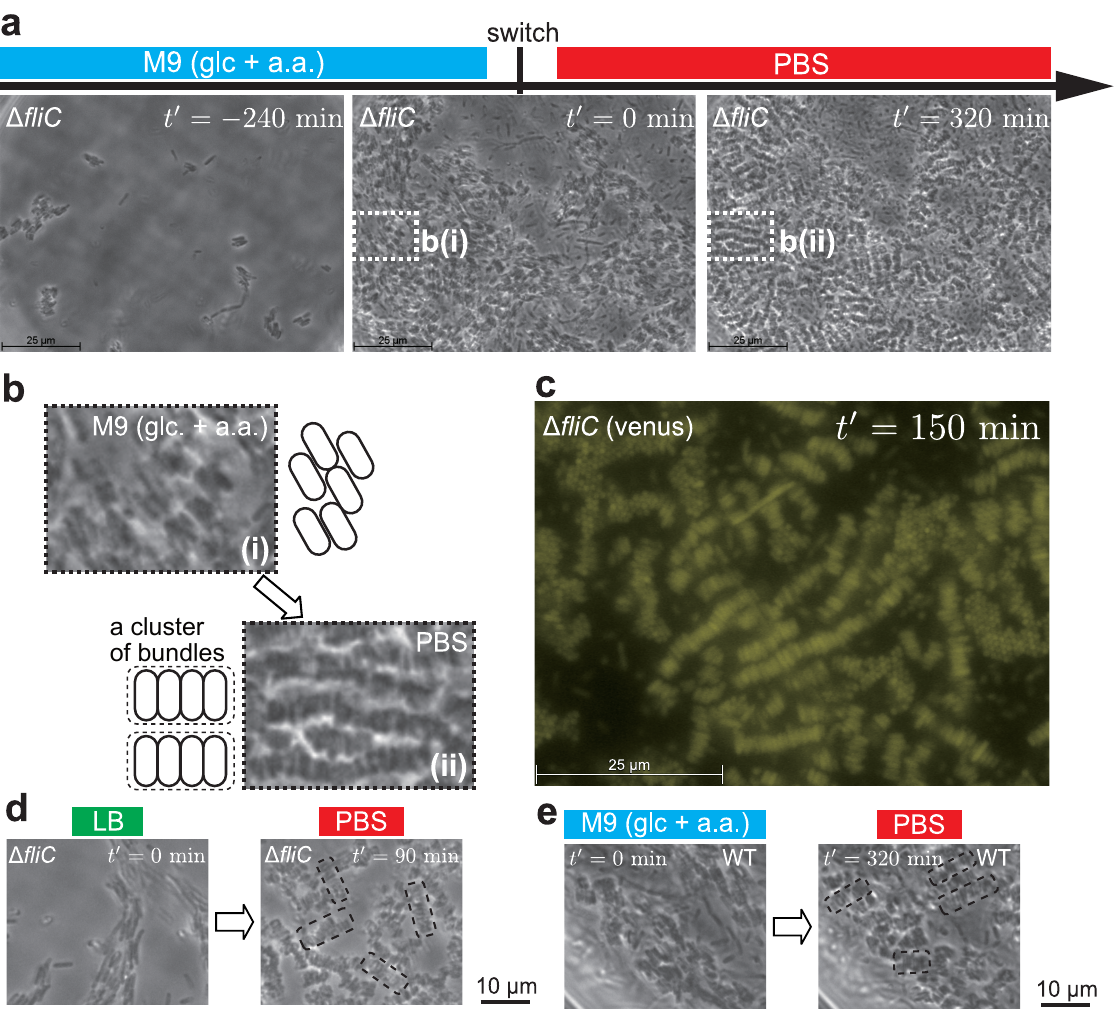}
       \caption{
       \label{fig-2}
       Bundle formation upon nutrient starvation.
       (a) Observation of the $\Delta$\textit{fliC} mutant grown in M9(glc+a.a.) and starved in PBS.
       $t'=0$ is the time at which the environment was switched.
       The well diameter $\phi$ is $150\unit{\mu{}m}$.
       Bundle clusters were observed in all 30 wells with various diameters $\phi$ recorded in a single experiment (``$\Delta$\textit{fliC} starvation'' in \tbref{tb-2}).
       See also \vidref4.
       (b) Enlargement of the regions outlined by the dotted rectangles in (a) at $t'=0\unit{min}$ (i) and $t'=320\unit{min}$ (ii).
       In (i), cells were nematically aligned. In (ii), these cells arranged themselves next to each other, and bundle clusters were formed.
       (c) A fluorescent image of bundles of the $\Delta$\textit{fliC} (\textit{venus}) mutant grown in M9(glc+a.a.) and starved in PBS. 
       %taken in another technical replicate ($\Delta$\textit{fliC} (\textit{venus}) starvation in \tbref{tb-2}) from the experiments shown in a and b.
       The well diameter $\phi$ is $330\unit{\mu{}m}$.
       (d) Observation of the $\Delta$\textit{fliC} mutant grown in LB and starved in PBS.
       The two snapshots were taken from the same position.
       Some regions with bundles are emphasized by dotted lines.
       The well diameter $\phi$ is $110\unit{\mu{}m}$.
       Bundle clusters were observed in 23 out of 30 wells with various diameters recorded in a single experiment (``$\Delta$\textit{fliC} starvation (LB)'' in \tbref{tb-2}).
       %, and not in the other wells.
       See also \vidref7.
       (e) Observation of the WT in the case of M9(glc+a.a.) $\to$ PBS.
       The two snapshots were taken from the same position.
       Some regions with bundles are emphasized by dotted lines.
       The well diameter $\phi$ is $150\unit{\mu{}m}$.
       Bundle clusters were observed in 15 out of 18 wells with various diameters recorded in a single experiment (``WT starvation'' in \tbref{tb-2}).
       %, and not in the other wells.
       See also \vidref8.
       }
\end{figure*}

\begin{figure*}[t]
       \includegraphics[width=\hsize]{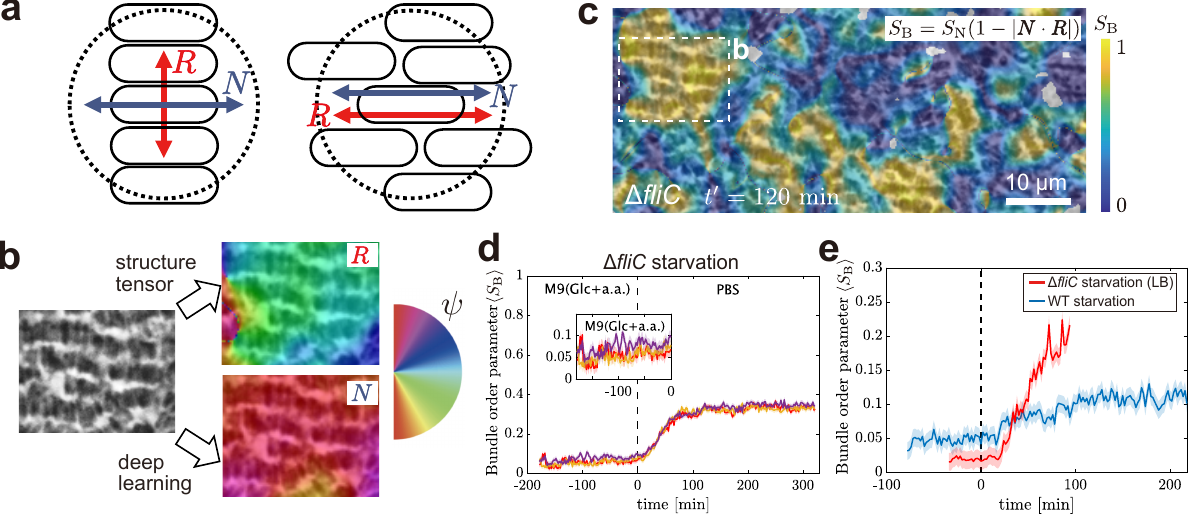}
       \caption{
       \label{fig-3}
       Quantitative evaluation of the bundle formation.
       (a) Illustrations of the definition of the density correlation director ($\bm{R}$) and the cell orientation ($\bm{N}$).
       (Left) When bundles are formed, $\bm{R}$ is oriented along the bundle axis, so that $\bm{R}$ and $\bm{N}$ are perpendicular.
       The dotted circle illustrates the approximate range of the Gaussian kernel used in \eqref{eq-1}.
       (Right) Without bundles, $\bm{R}$ is along the long axis of cells, parallel to $\bm{N}$.
       (b) Estimation of $\bm{R}$ and $\bm{N}$.
       The image is enlargement of the rectangle region in (c).
       The density correlation director $\bm{R}$ is obtained by the structure tensor method [\eqref{eq-1}].
       The cell orientation is obtained by the deep learning model which we trained in this study.
       See Materials and Methods and \figsref{S-figS2} and \ref{S-figS3}.
       (c) The evaluated bundle order parameter field $S_\mathrm{B}$.
       The image is from the $\Delta$\textit{fliC} starvation shown in \figref{fig-2}a,b.
       (d) The spatially-averaged bundle order parameter in the regions where aggregates exist, $\expct{S_\mathrm{B}}$, for the $\Delta$\textit{fliC} starvation.
       Each time series represents the data in a separate well ($\phi=150\unit{\mu{}m}$).
       The inset shows enlargement of the data before.
       The observation shown in b,c, \figref{fig-2}a,b, and \vidsref{4, 5, and 6} corresponds to the purple curve.
       The error is shown by the shaded bands, which was evaluated as the mean absolute error of the trained deep learning model compared to the ground truth by manual annotation (see Materials and Methods for details).
       (e) The spatially-averaged bundle order parameter $\expct{S_\mathrm{B}}$ for the $\Delta$\textit{fliC} starvation (LB) and the WT starvation. The former corresponds to the observation shown in \figref{fig-2}d and \vidref7, and the latter to \figref{fig-2}e and \vidref8.
       }
\end{figure*}

%Motivated by the above facts, 
We then observed how planktonic aggregates respond to nutrient starvation.
We first grew bacteria (the $\Delta$\textit{fliC} mutant) in a nutritious condition to produce aggregates, then replaced the growth medium with non-nutritious phosphate-buffered saline (PBS) while observing the aggregates (see Materials and Methods).
Then we found, as shown in \figref{fig-2}a,b and \vidref4,
%for the $\Delta$\textit{fliC} mutant grown in M9(glc+a.a.) and starved in PBS (denoted by $\Delta$\textit{fliC} starvation in \tbref{tb-2}),
the formation of bundle clusters, where multiple cells clumped together next to each other.
%with their centers of mass aligned in the short axis of the cell.
Such bundle clusters may also gather and develop a structure similar to that of the smectic A phase of liquid crystal (\figref{fig-2}b).
Bundle clusters can be observed more clearly by fluorescent microscopy (\figref{fig-2}c) by using a strain with a fluorescent protein Venus \cite{Nagai2002}.
% Some cells are verticalized and aggregate in a hexagonal lattice, suggesting that the bundle structure is formed three-dimensionally as well.
%To see whether this bundle formation is robust, we attempted to reproduce the bundle using different starvation conditions and strains.
Also, the bundle formation was not limited to the specific condition and strain chosen here.
We observed similar bundle clusters when we grew cells in LB medium (\figref{fig-2}d and \vidref7; ``$\Delta$\textit{fliC} starvation (LB)'' in \tbref{tb-2}) and when we used the WT expressing flagella (\figref{fig-2}e and \vidref8; ``WT starvation'' in \tbref{tb-2}).
This indicates the robustness of our finding of the bundle formation upon starvation, regardless of detailed conditions and cellular states.

To quantify the degree of the smectic order, we define the bundle order parameter $S_\mathrm{B}$ by
\begin{equation}
       S_\mathrm{B} = S_\mathrm{N}(1-|\bm{N}\cdot\bm{R}|).  \label{eq-3}
\end{equation}
Here, $\bm{N}$ denotes the coarse-grained field of the cell orientation, $\bm{R}$ the orientation of the density correlation, and $S_\mathrm{N}$ the nematic order parameter evaluated from the cell orientation $\bm{N}$ (\figref{fig-3}a).
The two orientations $\bm{R}$ and $\bm{N}$ were evaluated by the structure tensor method \cite{Jaehne-book1993} and by a supervised deep neural network method, U-Net \cite{Ronneberger2015}, respectively (see Materials and Methods for details).
The bundle order parameter $S_\mathrm{B}$ takes a large value when the nematic order is strong and $\bm{N}$ and $\bm{R}$ are nearly perpendicular.
The latter condition is set because, when bundle clusters are formed, bacterial cells arrange themselves next to each other, in the direction of their short axis (\figref{fig-3}b), similarly to the smectic A phase.
We also observe that, when no bundles are formed, cells are typically located along their long axis, resulting in stronger density correlation in that direction, $\bm{R} \parallel \bm{N}$.
Thus, our bundle order parameter $S_\mathrm{B}$ can evaluate the degree of bundle formation and its smectic order, as demonstrated in \figref{fig-3}c and \vidsref{5 and 6}.
In the past, a similar order parameter was adopted in Ref.~\cite{Bakker2016} to evaluate the smectic order of colloidal rods.
 
\begin{figure*}[t]
       \includegraphics[width=\hsize]{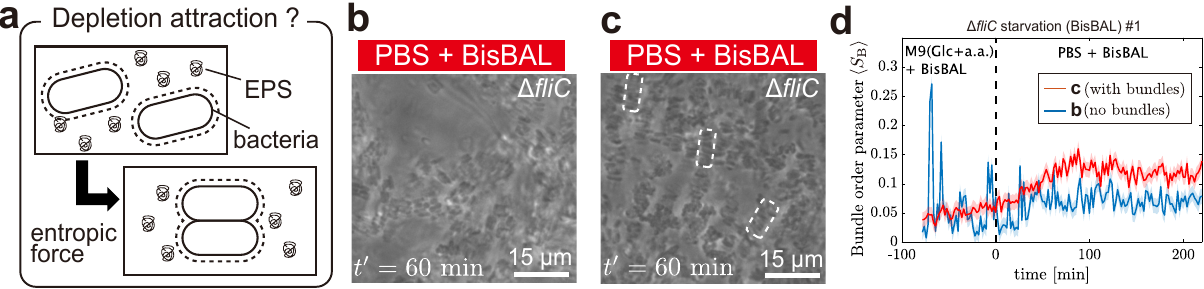}
       \caption{
       \label{fig-4}
       Investigation of the relevance of EPS.
       (a) Illustration of the hypothesis that the depletion attraction is relevant to the bundle formation.
       The dotted lines around the bacterial cells represent the excluded region which large molecules such as EPS cannot enter.
       Cells may form bundles to minimize the excluded volume and increase the entropy of large molecules.
       (b,c) Two examples of starvation response of the $\Delta$\textit{fliC} mutant grown in M9(glc+a.a.) + $2\unit{\mu{}M}$ BisBAL and starved in PBS + $2\unit{\mu{}M}$ BisBAL.
       The well diameter $\phi$ is $150\unit{\mu{}m}$.
       $t'=0$ is the time at which the environment was switched.
       The bundle formation was not observed at all (b and \vidref9) in 16 out of 30 wells with various diameters recorded in a single experiment ($\Delta$\textit{fliC} starvation (BisBAL) \#{}1 in \tbref{tb-2}), while it was observed (c and \vidref{10}) in the other wells.
       We also conducted a technical replicate ($\Delta$\textit{fliC} starvation (BisBAL) \#{}2 in \tbref{tb-2}) and found no bundle formation in 9 out of 13 wells.
       These indicate that comprehensive reduction of EPS secretion by BisBAL significantly inhibits the bundle formation, suggesting the possible relevance of depletion attraction mediated by EPS.
       (d) The spatially-averaged bundle order parameter $\expct{S_\mathrm{B}}$ for the cases where bundles were formed and not formed.
       }
\end{figure*}

To evaluate the bundle formation process, we measured the time evolution of the spatially-averaged bundle order parameter, $\expct{S_\mathrm{B}}$.
The spatial average was calculated over the regions where aggregates exist, which were also obtained by the deep neural network (see Materials and Methods).
We found that $\expct{S_\mathrm{B}}$ increases significantly after the starvation, in all cases described so far [the $\Delta$\textit{fliC} starvation (\figref{fig-3}d), the $\Delta$\textit{fliC} starvation (LB) and the WT starvation (\figref{fig-3}e)].

\begin{figure*}[t]
       \includegraphics[width=\hsize]{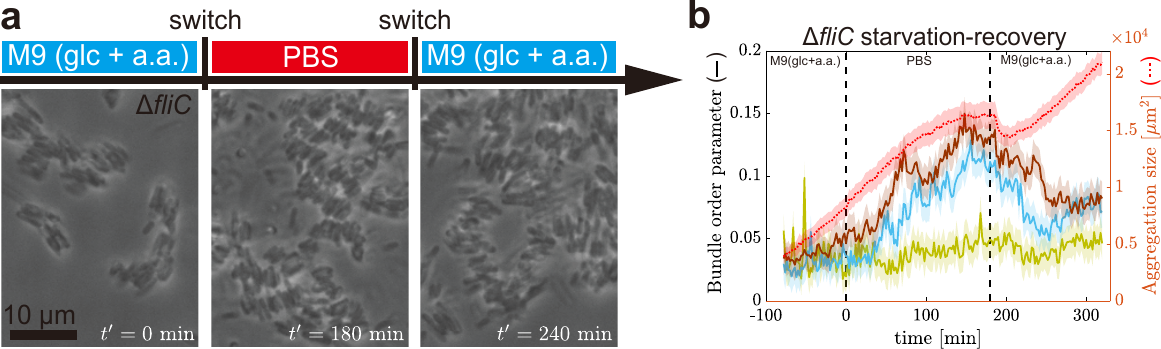}
       \caption{
       \label{fig-5}
       Investigation of the relevance of cellular growth arrest.
       (a) Observations of the $\Delta$\textit{fliC} mutant grown in M9(glc+a.a.), starved in PBS, then grown again in M9(glc+a.a.).
       $t'=0$ is the time at which the environment was switched to PBS, and the switch back to M9(glc+a.a.) was at $t'=180\unit{min}$.
       The well diameter $\phi$ is $150\unit{\mu{}m}$.
       Bundle clusters were observed in 25 out of 27 wells with various diameters recorded in a single experiment ($\Delta$\textit{fliC} starvation-recovery in \tbref{tb-2}).
       Those bundles were destroyed after the recovery of cell growth by M9(glc+a.a.) in all wells.
       The displayed snapshots were taken from the same position.
       See also \vidref{15}.
       (b) The spatially-averaged bundle order parameter over the region where aggregates exist, $\expct{S_\mathrm{B}}$, in the $\Delta$\textit{fliC} starvation-recovery experiment.
       Each time series represents the data in a separate well ($\phi=230\unit{\mu{}m}$).
       The observation shown in (a) and \vidref{15} corresponds to the blue curve.
       The definition of the errors (shades) is the same as that for \figref{fig-3}d.
       %The error bar was evaluated using the bootstrap method (see Materials and Methods for details), which was smaller than the symbol size.
       We also plot the aggregation size, summed over the three wells (red dotted line).
       The displayed error for the aggregation size is the sum of the errors of the individual aggregate sizes. The error of each aggregate size was estimated to be the aggregate perimeter $\times$ 5 pixels ($\sim 0.85\unit{\mu{}m}$), which we consider to be the range of uncertainty in the edge location of the segmented region.
       The sudden decrease of the total aggregation size may be because of a weak hydrodynamic disturbance caused by the medium switch.
       }
\end{figure*}

We should note, however, that the value of the bundle order parameter is sensitive to subtle differences in experimental conditions, such as the weakly heterogeneous image background due to the porous membrane.
For example, in another measurement of the $\Delta$\textit{fliC} mutant strain from M9(glc.+a.a.) to PBS (specifically ``$\Delta$\textit{fliC} starvation-recovery'' in \tbref{tb-2}, to be described below), the bundle order parameter during the starvation (\figref{fig-5}b) was significantly smaller than that of the $\Delta$\textit{fliC} starvation (\figref{fig-3}d).
Therefore, we compare the values of the bundle order parameter only within each technical replicate, to characterize the structure before and after starvation.

\subsection{The relevance of extracellular molecules}

Here we consider what changes in cell-cell interactions occur upon starvation and result in the bundle formation.
In the literature, qualitatively similar bundles of rod-shaped cells have been observed in the presence of a sufficient amount of polymer \cite{Dorken2012,Secor2018,Azimi2021,LoboCabrera2021}.
This was interpreted as a result of depletion attraction between cells.
A similar effect has also been reported for mixtures of rod-shaped and spherical colloids \cite{Koda1996,Adams1998a,Adams1998b,Ye2013,Bakker2016}, resulting in the smectic A phase similar to our observation.
Since the secretion of some EPS is known to be promoted under external stresses \cite{Myszka2009,Myszka2011,Zhang2014,Costa2018}, we hypothesize that the EPS-induced depletion attraction may be the mechanism of the bundle formation upon starvation (\figref{fig-4}a).

To test this hypothesis, we observed the starvation response of aggregates in the presence of a reagent that comprehensively inhibits EPS secretion.
Specifically, we used bismuth dimercaprol (BisBAL), which has been reported to comprehensively reduce EPS secretion roughly by a factor of 10 under appropriate concentrations, without significantly affecting cell growth \cite{Badireddy2008a,Badireddy2008b}.
We cultured, in the presence of BisBAL, populations of the $\Delta$\textit{fliC} mutant forming aggregates in M9(Glc+a.a.) and exposed them to starvation in PBS (``$\Delta$\textit{fliC} starvation (BisBAL)'' in \tbref{tb-2}).
In contrast to the case without BisBAL, where bundles were formed in all monitored wells ($\Delta$\textit{fliC} starvation), with BisBAL we observed that bundles were not formed (\figref{fig-4}b and \vidref9) in more than half of the observed wells in two technical replicates (\#{}1 and \#{}2 in \tbref{tb-2}), while in the other wells bundles were formed relatively weakly (\figref{fig-4}c and \vidref{10}); see also the bundle order parameter for those two cases (\figref{fig-4}d).
This indicates that the reduction of EPS secretion by BisBAL significantly, albeit not completely, inhibits the bundle formation.
Our results suggest the relevance of depletion attraction by EPS.

While the comprehensive inhibition of EPS secretion by BisBAL allowed us to know that EPS is involved in the bundle formation process, it is useful to identify specific substances that serve as depletants of this depletion attraction.
There are several candidate substances, whose secretion is known to increase during starvation and whose size is large enough to cause depletion attraction between cells.
First, \textit{E. coli} is known to promote the secretion of cellulose under environmental stresses \cite{Gualdi2008,Serra2013}. 
However, we can immediately rule out its relevance because the W3110 strain used in this study does not produce cellulose \cite{Serra2013}.
Second, curli, polymeric amyloid fibers which usually extend from and attach to cell surfaces \cite{Barnhart2006}, is known to be produced more actively under environment stresses \cite{Brombacher2006,Barnhart2006}.
It was reported that its major subunit protein, CsgA, can polymerize in medium being detached from cell surfaces \cite{Wang2007}, which may induce depletion attraction.
Third, outer membrane vesicles (OMVs) have a diameter of $20$--$250\unit{nm}$, which contain, e.g., misfolded proteins and toxic substances, and are discharged from cells \cite{Schwechheimer2015}.
It has been reported that the secretion of OMVs is promoted under external stress \cite{Schwechheimer2015,Roier2016} and that they can induce depletion attraction to particles on the scale of tens of micrometers \cite{Amimanan2017,Kanlaya2019}.
Therefore, we tested the possible influence of curli and OMVs on the bundle formation.

We first tested the possibility of curli, by using a mutant W3110 $\Delta$\textit{fliC} \textit{csgA}::\textit{kan} (``$\Delta$\textit{fliC} \textit{csgA}::\textit{kan}'' mutant in \tbref{tb-1}) for which the gene expressing CsgA protein
%, a monomer that constitutes curli, 
was replaced by a kanamycin-resistant gene \textit{kan}.
On an agar plate of LB broth containing congo red, which selectively stains amyloid fibers including curli \cite{Reichhardt2015}, we confirmed that the $\Delta$\textit{fliC} \textit{csgA}::\textit{kan} mutant does not produce curli (\supfigref{S-figS4}a).
We then observed aggregates of the mutant grown in M9 (glc+a.a.) and starved in PBS (``$\Delta$\textit{fliC} \textit{csgA}::\textit{kan} starvation'' in \tbref{tb-2}) and found bundle clusters in most wells (\tbref{tb-2}, \supfigref{S-figS4}b and \vidref{11}).
These indicate that curli may not be the main depetant, at least not the only one, for the bundle formation upon starvation that we observe.

Next, we investigated the influence of OMVs.
If OVMs are main depletants, a significant amount of OVMs need to be released in the medium.
Since OVMs are produced by cells, their density is expected to be higher inside bundle clusters, between individual bundles (region between dashed rectangles in the sketch of \figref{fig-2}b(ii)).
We attempted to inspect this, by staining cell membranes and OVMs with FM 4-64 dye, which selectively label hydrophobic cell membranes including OMVs \cite{Toyofuku2017,Mandal2021} (\supfigref{S-figS5}).
By using the W3110 $\Delta$\textit{fliC} \textit{intS}::\textit{venus}-\textit{kan} mutant (``$\Delta$\textit{fliC} (\textit{venus}) mutant'' in \tbref{tb-1}), the cell bodies were also visualized by the fluorescent protein Venus.
We observed aggregates of this strain grown in M9(glc.+a.a.) and starved in PBS (``$\Delta$\textit{fliC} (\textit{venus}) starvation'' in \tbref{tb-2}).
However, the results in \supfigref{S-figS5} did not show an appreciable amount of OVMs in the medium, even between individual bundles inside clusters (compare the fluorescence intensities of Venus and FM 4-64).
This suggests that OMVs may not be the main depletant that triggers the bundle formation.

As another test, we attempted to induce bundle formation by promoting the secretion of OMVs by antibiotics, instead of starvation.
Specifically, we used polymyxin B, known to increase the OMV secretion by several times to tenfold at sub-minimum inhibitory concentrations (sub-MIC) \cite{Manning2011}.
We cultured the $\Delta$\textit{fliC} mutant in growth medium (either M9(glc.+a.a.) or LB) containing sub-MIC polymyxin B for long time, without starvation, but no bundle clusters were observed at all (``$\Delta$\textit{fliC} growth (PMB)'' and ``$\Delta$\textit{fliC} growth (LB, PMB)'' in \tbref{tb-2}, \supfigref{S-figS6}a,b and \vidsref{12 and 13}).

We also attempted to induce the bundle formation by an external addition of a standard depletant, namely xanthan gum, an extracellular polysaccharide produced by \textit{Xanthomonas campestris} \cite{Palaniraj2011}, which was used in the literature to introduce depletion attraction between cells \cite{Dorken2012}.
Following Refs.~\cite{Khan2018,Yang2019}, we supplemented xanthan gum to medium, at a concentration sufficient to reproduce the viscosity due to EPS secreted by \textit{E. coli}.
We observed aggregates of the $\Delta$\textit{fliC} mutant grown in M9(glc.+a.a.) (``$\Delta$\textit{fliC} growth (xanthan)'' in \tbref{tb-2}).
In the early stage while the cell density was low, we observed a few small bundles containing 2-3 cells that seem to be induced by the depletion attraction (see \vidref{14}).
However, the small bundles immediately disappeared afterward, and no bundle formation was finally realized in any of the observed wells (\tbref{tb-2} and \supfigref{S-figS6}c).
An interpretation of these observations under growth conditions with depletants will be given in the next subsection.

Our series of experiments suggest that, while depletion attraction via EPS seems to contribute significantly to the bundle formation upon starvation, the presence of depletant solely is not sufficient to induce the bundle formation.
It is in line with a recent inspection of bacterial aggregate patterns under controlled polymeric environments \cite{LaCorte.etal-b2024}, where no smectic-like bundle was observed for any polymer conditions studied therein.

\subsection{The relevance of growth arrest}

Now we consider the possibility that the arrest of cell growth by nutrient starvation may also have a significant effect on the bundle formation.
We hypothesize that, even if a sufficient amount of depletants are diffused between cells, perturbations due to cell elongation and division may overcome the depletion attraction and prevent bundle formation.

To test this hypothesis, we attempted to switch back from the non-nutritious to nutritious conditions, after bundle clusters were formed upon starvation. 
Specifically, we first prepared bundle clusters of the $\Delta$\textit{fliC}, grown in M9(glc.+a.a.) and starved in PBS, as in the $\Delta$\textit{fliC} starvation, then resumed cell growth by switching it back to M9(glc.+a.a.) (``$\Delta$\textit{fliC} starvation-recovery'' in \tbref{tb-2}; \figref{fig-5}a and \vidref{15}).
We measured the time evolution of the bundle order parameter in three independent wells, as well as the total size of aggregates in those wells (\figref{fig-5}b).
During the starvation stage (PBS), the bundle order parameter increased, and the total aggregate size slows down after a lag time (red dotted line).
Then, after the growth condition was recovered, the aggregate size started to grow again, while the smectic-like bundles were destroyed and the bundle order parameter decreased.
Therefore, not only the depletion attraction induced by EPS but also growth arrest is necessary for the bundle formation we observed.
This may also be the reason why bundle clusters were not formed in the aforementioned $\Delta$\textit{fliC} growth (PMB), $\Delta$\textit{fliC} growth (LB, PMB), and $\Delta$\textit{fliC} growth (xanthan) experiments, where an increased amount of depletants was present but growth was not arrested.
This may also explain why bundle clusters were not observed in Ref.\,\cite{LaCorte.etal-b2024}.

\section{Conclusions}

In this study, using the EMPS we developed previously, we realized observations of planktonic aggregates of non-motile \textit{E. coli} under time-dependent starving conditions.
% Observations in a steady environment showed that the size of aggregates increased exponentially, and their growth rate was comparable to that of the single cells.
We discovered that nutrient starvation altered the structure of aggregates, resulting in the appearance of smectic-like bundle clusters, where cells arranged themselves next to each other along their short axis.
By utilizing the deep learning method, we successfully quantified the degree of smectic order of the bundle clusters.
We revealed that EPS produced by cells is involved in the bundle formation, by showing that comprehensive reduction of EPS secretion prevented the bundle formation.
%We also inspected the effects of several specific substances, curli, OMVs, and xanthan gum.
We also found that the smectic-like bundles were destroyed by resumption of growth after recovery of nutrients.
Our results indicate that both the depletion attraction via EPS and growth arrest of cells are relevant to the bundle formation we observed.

Further investigation is required for elucidating what specific substances are most relevant to the depletion attraction involved.
In this work, we inspected two specific substances, curli and OMVs, but could not determine their relevance conclusively.
We also remark that it is probably difficult to validate the relevance of OMVs by inhibition, because complete inhibition of OMVs seems to be unrealistic for living cells. 
It is worthwhile to visualize spatial distribution of OMVs in the presence of bundle clusters, by selectively labeling OMVs with, e.g., curvature-sensing peptides \cite{Kawano2020}.
In the literature, it is also known that changes in cell surface conditions may also affect the formation of similar bundle clusters.
For \textit{P. aeruginosa} populations, it was observed that decreasing the hydrophobicity of O-specific antigen on the cell surface resulted in the reduction of the hydrophobic attraction between cells, which facilitated the bundle formation by polymers added in the environment \cite{Azimi2021}.
In the case of \textit{E. coli}, it was reported that cells in the stationary phase develop heterogeneous charge distribution on the outer membranes, reducing the electrostatic repulsions between cells and surfaces \cite{Walker2005}.
It may also be important to seek for the possible existence of unidentified processes underlying the bundle formation we observed.
We also note the apparent similarity to the rouleaux formation of red blood cells \cite{WAGNER2013}.

The physiological significance of the bundle formation needs to be investigated in future.
For the case of the nematic ordering of anisotropic cells, previous studies \cite{Hrabe2004,Cho2007} proposed its effect on the diffusion efficiency of small molecules, such as nutrients and signal molecules produced by cells, in the intercellular gaps.
Specifically, in the case of dense \textit{E. coli} populations, it has been experimentally suggested that the transport efficiency along the director of the nematic order can be enhanced \cite{Cho2007}.
%The efficient transportation was attributed to the reduction in the effective path length of the intercellular space along the nematic director.
It may be interesting to consider an analogous effect in the presence of the smectic order as we observed in this work.
If the bundle formation is beneficial for, e.g., an uptake of nutrients from the external environment under starvation, our finding may be a behavioral adaptation of bacterial populations.
Besides, the biological functions of liquid crystalline order -- mostly nematic -- in cell populations are currently under active investigation \cite{Doostmohammadi.etal-NC2018,Doostmohammadi.Ladoux-TCB2022}, including bacterial populations \cite{DellArciprete.etal-NC2018,Copenhagen.etal-NP2021,Shimaya.Takeuchi-PN2022,Nijjer.etal-NP2023}.
Our finding of the smectic order in bacteria may suggest the potential importance of this liquid crystalline state too, calling for further investigations of its relevance from various perspectives.

Finally, our findings may also be relevant to other bacteria, because the elevated EPS secretion and the growth arrest by starvation are not unique to \textit{E. coli}.
Further elucidation of the mechanism of the bundle formation and its biological significance is an important future task, which may also contribute to understanding the physical mechanisms underlying biofilm development.

\section{Materials and methods}

\begin{table*}[t]\centering
       \caption{\textit{E. coli} strains used in this study.}
       \label{tb-1}
       \begin{tabular}{l|l|l}
          Strains & Genotypes & Characteristics \tabularnewline
          \hline
          WT & The wild-type of W3110 & \tabularnewline
          \textit{flu}::\textit{kan} mutant & W3110 \textit{flu}::\textit{kan} & no Ag43 \tabularnewline
          $\Delta$\textit{fliC} mutant & W3110 $\Delta$\textit{fliC} & no flagella \tabularnewline
          $\Delta$\textit{fliC} \textit{csgA}::\textit{kan} mutant & W3110 $\Delta$\textit{fliC} \textit{csgA}::\textit{kan} & no flagella nor curli \tabularnewline
          $\Delta$\textit{fliC} (\textit{venus}) mutant & W3110 $\Delta$\textit{fliC} \textit{intS}::\textit{venus}-\textit{kan} & no flagella, expressing the fluorescent protein Venus \tabularnewline
          \hline
       \end{tabular}
\end{table*}

\begin{table*}[t]\centering
       \caption{
       List of measurements in this study.
       The column ``Bundle fraction'' indicates the fraction of the wells where the bundle formation was observed, showing the number of such wells as the numerator and the total number of the monitored wells as the denominator.}
       \label{tb-2}
       \begin{tabular}{l|l|l|c|l}
              Measurement & Strains & Conditions & Bundle fraction & Note \tabularnewline
              \hline\hline
              WT growth & WT & M9(glc+a.a.) & & \supfigref{S-figS1}a, \vidref1 \tabularnewline \hline
              \textit{flu}::\textit{kan} growth & \textit{flu}::\textit{kan}  & M9(glc+a.a.) & & \supfigref{S-figS1}b, \vidref2 \tabularnewline \hline
              $\Delta$\textit{fliC} growth & $\Delta$\textit{fliC}  & M9(glc+a.a.) & & \supfigref{S-figS1}c, \vidref3 \tabularnewline \hline
              $\Delta$\textit{fliC} starvation & $\Delta$\textit{fliC}  & M9(glc+a.a.) $\to$ PBS & 30/30 & \figref{fig-2}a,b, \vidref4 \tabularnewline \hline
              $\Delta$\textit{fliC} (\textit{venus}) starvation & $\Delta$\textit{fliC} (\textit{venus})  & M9(glc+a.a.) $\to$ PBS & & \figref{fig-2}c, \supfigref{S-figS5} \tabularnewline
               & & & & for fluorescence imaging \tabularnewline \hline
              $\Delta$\textit{fliC} starvation (LB) & $\Delta$\textit{fliC}  & LB $\to$ PBS & 23/30 & \figref{fig-2}d, \vidref7 \tabularnewline \hline
              WT starvation & WT & M9(glc+a.a.) $\to$ PBS & 15/18 & \figref{fig-2}e, \vidref8 \tabularnewline
               & & & & expressing flagella \tabularnewline \hline
              $\Delta$\textit{fliC} starvation (BisBAL) & $\Delta$\textit{fliC}  & M9(glc+a.a.) + 2$\unit{\mu{}M}$ BisBAL & 14/30 & \figref{fig-4}, \vidsref{9 and 10} \tabularnewline
              \#{}1 & & $\to$ PBS + 2$\unit{\mu{}M}$ BisBAL &  & reducing EPS production \tabularnewline \hline              
              $\Delta$\textit{fliC} starvation (BisBAL) & $\Delta$\textit{fliC}  & M9(glc+a.a.) + 2$\unit{\mu{}M}$ BisBAL & 4/13 & technical replicate \tabularnewline
              \#{}2 & & $\to$ PBS + 2$\unit{\mu{}M}$ BisBAL &  &  \tabularnewline \hline
              $\Delta$\textit{fliC} \textit{csgA}::\textit{kan} starvation & $\Delta$\textit{fliC} \textit{csgA}::\textit{kan}  & M9(glc+a.a.) $\to$ PBS & 15/19 & \supfigref{S-figS4}, \vidref{11} \tabularnewline 
               & & & & deleting curli \tabularnewline \hline
              $\Delta$\textit{fliC} growth (PMB) & $\Delta$\textit{fliC}  & M9(glc+a.a.) & 0/6 & \supfigref{S-figS6}a, \vidref{12} \tabularnewline
               & & + 250 ng/ml polymyxin B &  & promoting OMVs production \tabularnewline \hline
              $\Delta$\textit{fliC} growth (LB, PMB) & $\Delta$\textit{fliC}  & LB + 250 ng/ml polymyxin B & 0/15 & \supfigref{S-figS6}b, \vidref{13} \tabularnewline
               & & & & promoting OMVs production \tabularnewline \hline
              $\Delta$\textit{fliC} growth (xanthan) & $\Delta$\textit{fliC}  & M9(Glc+a.a.) & 0/6 & \supfigref{S-figS6}c, \vidref{14} \tabularnewline
              & & + 0.03 wt\% xanthan gum  &  \tabularnewline \hline
              $\Delta$\textit{fliC} starvation-recovery & $\Delta$\textit{fliC}  & M9(Glc+a.a.) & 25/27 & \figref{fig-5}, \vidref{15} \tabularnewline
              & & $\to$ PBS $\to$ M9(Glc+a.a.) &  \tabularnewline \hline
       \end{tabular}
\end{table*}

\subsection{Strains and culture media}

We used a wild-type (WT) \textit{E. coli} strain W3110 and mutant strains W3110 $\Delta$\textit{fliC}, W3110 \textit{flu}::\textit{kan}, W3110 $\Delta$\textit{fliC} \textit{csgA}::\textit{kan}, and W3110 $\Delta$\textit{fliC} \textit{insS}::\textit{venus}-\textit{kan} (see \tbref{tb-1} for the strain names and their genotypes).
We used the $\Delta$\textit{fliC} mutant prepared in Ref.~\cite{Hashimoto2016}.
By P1 transduction, the genes \textit{flu}::\textit{kan} and \textit{csgA}::\textit{kan} were transferred to the WT and the $\Delta$\textit{fliC} mutant, respectively, from the Keio collection of \textit{E. coli} through National BioResource Project (NBRP) of National Institute of Genetics, Japan.
Concerning the strain $\Delta$\textit{fliC} (\textit{venus}) mutant, we used a strain that was previously prepared by a method similar to that in Ref.~\cite{Koganezawa2021}.
The protein Venus was developed in Ref.~\cite{Nagai2002} and provided by the RIKEN BRC through NBRP of the MEXT/AMED, Japan.

\tbref{tb-2} is a list of the experiments conducted in this study.
The media we mainly used were M9(glc.+a.a.) (glucose 0.2 wt\% and MEM Amino Acids solution (M5550, Sigma) 1 wt\% in the M9 medium), LB broth (tryptone 1 wt\%, sodium chloride 1 wt\% and Yeast extract 0.5 wt\%), and the phosphate-buffered saline (PBS).
%2$\unit{\mu{}M}$ BisBAL, 250 ng/ml polymyxin B (PMB), and 0.03 wt\% xanthan gum were also supplemented to the medium in several experiments.
%Details on the culture condition used in each experiment are provided in \tbref{tb-2}.

\subsection{Evaluation of the loss of curli by congo red}

We verified the loss of curli in the $\Delta$\textit{fliC} \textit{csgA}::\textit{kan} mutant as follows.
We first inoculated the strain from a glycerol stock into $2\unit{ml}$ LB in a test tube and shook it overnight at $37\degc$.
We then inoculated $10\unit{\mu{}l}$ of the incubated suspension to an LB agar containing $25\unit{\mu{}g/ml}$ of congo red, by an inoculation loop.
We then obtained colony images of the $\Delta$\textit{fliC} mutant and the $\Delta$\textit{fliC} \textit{csgA}::\textit{kan} mutant, confirming the absence of dye fluorescence from the latter mutant, due to the loss of curli (\supfigref{S-figS4}a).

\subsection{Time-lapse observations}

Time-lapse observations were conducted by using the extensive microperfusion system (EMPS) previously developed by us and coworkers \cite{Shimaya2021}.
We fabricated $11\unit{\mu{}m}$ deep circular wells with varying diameter $\phi=110, 150, 230, 330\unit{\mu{}m}$ on a coverslip (\figref{fig-1}a).
After introducing cell suspension, these wells were covered by a rigid porous membrane, attached by biotin-streptavidin bonding. 
The porous membrane is a bilayer of a streptavidin-decorated cellulose membrane (Spectra/Por 7, Repligen, Waltham, MA, molecular weight cut-off 25,000) and a biotin-decorated PET membrane (taken from Transwell 3450, Corning, NY, nominal pore size $0.4\unit{\mu m}$).
See Ref.~\cite{Shimaya2021} for more details on the fabrication process of the EMPS.
%onto the substrate, we can trap bacterial populations in those wells where fresh medium is uniformly and constantly supplied through the porous membrane.

To prepare bacterial suspensions, we first inoculated the strain from a glycerol stock into $2\unit{ml}$ fresh medium in a test tube and shook it overnight at $37\degc$.
%(see \tbref{tb-2} for the choice of the medium for each experiment).
We used either M9(glc+a.a.) or LB broth, chosen accordingly to the growth medium used in the main experiment (see \tbref{tb-2}).
We then transferred $20\unit{\mu{}l}$ of the incubated suspension to $2\unit{ml}$ fresh medium (the same medium as that used in the main observation) and cultured it until the optical density (OD) at $600\unit{nm}$ wavelength reached $0.1$--$0.5$.
The cell suspension was finally diluted to $\mathrm{OD}=0.05$ before the inoculation on the substrate.

The device was placed on the microscope stage, in the incubation box maintained at $37\degc$ (\figref{fig-1}b).
We used a Leica DMi8 microscope equipped with a 63x (N.A. 1.30) oil immersion objective lens and operated by Leica LasX.
A syringe filled with the nutritious medium was connected to the device, and another syringe filled with non-nutritious buffer was also connected via a three-way junction with check valves when observing the starvation response.
Each syringe was set to a syringe pump (NE-1000, New Era Pump Systems).
To fill the device with the nutritious medium, we first injected the medium stored at $37\degc$ at the rate of $60\unit{ml/hr}$ for $5\unit{min}$. 
During the observation, first the nutritious medium was constantly supplied at the rate of $2\unit{ml/hr}$ (flow speed approximately $0.2\unit{mm/sec}$ above the membrane).
When changing the environment, the syringe pump for the nutritious medium was stopped, and the other pump for the non-nutritious buffer was turned on to replace the medium inside the device.
The flow rate was set to be $60\unit{ml/hr}$ for the first $5\unit{min}$, then returned to $2\unit{ml/hr}$.

Throughout the experiment, the device was always in the microscope incubation box, maintained at $37\degc$.
Cells were observed by phase-contrast microscopy focusing near the bottom of the wells and recorded at the time interval of $2\unit{min}$.
Images were taken from several wells (stated in the figure captions and \tbref{tb-2}) in each experiment.
The fraction of the wells showing bundle clusters presented in \tbref{tb-2} is based on visual assessment.

\subsection{Fluorescence labeling of cellular membranes by FM 4-64}

For the observation shown in \figref{fig-1}c and \supfigref{S-figS5}, we first inoculated the strain from a glycerol stock into $2\unit{ml}$ M9(Glc+a.a.) in a test tube and shook it overnight at $37\degc$.
We transferred $20\unit{\mu{}l}$ of the incubated suspension to $2\unit{ml}$ fresh M9(Glc+a.a.) and cultured it until OD reached $0.1$--$0.5$.
The cell suspension was finally diluted to $\mathrm{OD}=0.05$ before the inoculation on the device substrate.

Preparation of the device and the microscope, and induction of nutrient starvation were conducted similarly as for the time-lapse observations.
$150\unit{min}$ after the onset of starvation, when smectic-like bundle clusters appeared, PBS containing $5\unit{\mu{}g/ml}$ of FM 4-64 was introduced.
The medium flow was set to be $60\unit{ml/hr}$ for the first $5\unit{min}$ and left for $10\unit{min}$ without flow to stain cellular membranes.
We then washed unbound FM 4-64 by supplying PBS without FM 4-64 at the rate of $60\unit{ml/hr}$ for $3\unit{min}$, stopped the flow, and obtained fluorescence images.

\subsection{Image analysis}

We employed image analysis to evaluate the smectic order of bundle clusters, in terms of the bundle order parameter $S_\mathrm{B}$ [\eqref{eq-3}] as follows.
The entire pipeline of the image analysis is illustrated in \supfigref{S-figS2}.

First we detected the regions where planktonic aggregates existed.
We utilized the deep neural network, U-Net \cite{Ronneberger2015}, to automate this process.
For training and evaluating the model, we extracted 144 frames from all experiments at approximately constant time intervals.
Each image consists of $648 \times 483$ pixels, with pixel size $\approx 0.18\unit{\mu{}m}$. 
To prepare the teacher label, the target region was manually annotated by using painting software.
We also performed data augmentation through the horizontal and/or vertical flip and separated the data into five groups for cross-validation.
After normalizing the image intensity distribution by contrast limited adaptive histogram equalization (CLAHE), U-Net was trained using the binary cross entropy loss.
The segmentation accuracy of the trained model, evaluated by the Dice similarity coefficient (DSC), was $0.577$.

We then detected the two orientation fields, namely the density correlation director $\bm{R}$ and the cell orientation $\bm{N}$, separately.
For $\bm{R}$, we employed the structure tensor method \cite{Jaehne-book1993}. 
We calculated the structure tensor $J(\bm{r})$ at a given pixel $\bm{r}=(x,y)$ by
\begin{equation}
J(\bm{r}) = 
       \begin{pmatrix}
         [\Delta_x I,\Delta_x I]_{\bm{r}}, & [\Delta_y I,\Delta_x I]_{\bm{r}}  \\ \relax
         [\Delta_x I,\Delta_y I]_{\bm{r}}, & [\Delta_y I,\Delta_y I]_{\bm{r}} 
       \end{pmatrix}
     ,  \label{eq-1}
\end{equation}
with the image intensity $I(x,y)$, $\Delta_x I \equiv I(x+1,y)-I(x-1,y)$, $\Delta_y I \equiv I(x,y+1)-I(x,y-1)$, and $[g,h]_{\bm{r}}\equiv\sum_{(x',y')\in\mathrm{ROI}^\ell_{\bm{r}}}\ g(x',y')h(x',y')f_{\bm{r}}^\sigma(x',y')$.
The summation is taken over the region of interest ROI${}^\ell_{\bm{r}}$, which is a square of size $\ell \approx 7.2\unit{\mu{}m}$ ($40\unit{pixels}$) centered at $\bm{r}$, and $f_{\bm{r}}^\sigma(x',y')$ is the Gaussian kernel defined by $f_{\bm{r}}^\sigma(x',y')\equiv\exp\left[-\frac{(x'-x)^2+(y'-y)^2}{2\sigma^2}\right]$ with $\sigma \approx 1.8\unit{\mu{}m}$ ($10\unit{pixels}$).
Then $\bm{R}(\bm{r})$ is given by the eigenvector of $J(\bm{r})$ associated with the smallest eigenvalue.

For the cell orientation $\bm{N}$, we employed the U-Net again.
To obtain the teacher label, we manually annotated the long axis of cells by painting black straight lines (see \supfigref{S-figS2}).
We then calculated the structure tensor from the painted images to acquire the ground truth of the cell orientation, $\bm{\hat{N}}=\pm(\cos\hat{\psi}, \sin\hat{\psi})$ with $(-\pi/2\le \hat{\psi} < \pi/2)$.
The U-Net was trained to predict $\bm{N}=\pm(\cos\psi, \sin\psi)$ for each pixel in the CLAHE processed image, with the following nematic L2 loss function:
%defined as the following nematic L2 loss:
\begin{equation}
\ell_\mathrm{NL2}(\psi,\hat{\psi}) = \min(|\psi-\hat{\psi}|,|\pi-|\psi-\hat{\psi}||)^2
     .  \label{eq-4}
\end{equation}
During training, the loss function was evaluated only in the regions with planktonic aggregates obtained by the segmentation model.
The nematic L2 loss was then spatially averaged over all regions with planktonic cells, and the mini batch training with batch size $8$ was carried out.
The angle measurement accuracy of the trained model, evaluated by the mean absolute error of $\psi$, was $\pm 20.9 ^\circ$.
Using the predicted $\bm{N}$, we also evaluated the nematic order parameter by
\begin{equation}
       S_\mathrm{N}(\bm{r})=\expct{\sin2\psi}^2_{\mathrm{ROI}^\ell_{\bm{r}}} + \expct{\cos2\psi}^2_{\mathrm{ROI}^\ell_{\bm{r}}},  \label{eq-2}
\end{equation}
where $\expct{\cdot}_{\mathrm{ROI}^\ell_{\bm{r}}}$ denotes the spatial average within ROI${}^\ell_{\bm{r}}$.

Finally, using all these data, we calculated the bundle order parameter $S_\mathrm{B}$ [\eqref{eq-3}] for each pixel (see also \supfigref{S-figS2}).
We obtained the spatially-averaged bundle order parameter $\expct{S_\mathrm{B}}$ over the regions with planktonic aggregates, detected by the segmentation model.
The mean absolute error of $\expct{S_\mathrm{B}}$ per image was $\pm 0.013$ and shown as the shaded bands in the figures.

\section*{Author contributions}
T.S. and K.A.T. designed research.
T.S. performed all bacterial experiments and analyzed data.
T.S. and K.A.T. discussed the results and wrote the manuscript.

\section*{Conflicts of interest}
There are no conflicts to declare.

\section*{Acknowledgements}
We are grateful to F. Yokoyama for useful discussions that motivated us to investigate the influence of OMVs.
We thank Y. Wakamoto and R. Okura for sharing the P1 phage to prepare the \textit{flu::kan} mutant.
We also acknowledge discussions with S. Datta, D. Nishiguchi, Y. Okada, and M. Yanagisawa.
This work is supported in part by KAKENHI from Japan Society for the Promotion of Science (JSPS) (No. JP19H05800, JP20H00128, JP24K00593), KAKENHI for JSPS Fellows (No. JP20J10682), and the JSPS Core-to-Core Program ``Advanced core-to-core network for the physics of self-organizing active matter (JPJSCCA20230002)''

\section*{Data availability}
The code and data that support the findings of this study will be available at Zenodo.

\bibliography{./BDpaper}

\end{document}

\title{Supplementary Material for\\``Smectic-like bundle formation of planktonic bacteria upon nutrient starvation''}% Force line breaks with \\
%\title{Supplementary Material for\\``Bundle formation in planktonic \textit{E. coli} populations under nutrient starvation''}% Force line breaks with \\

\author{Takuro Shimaya}
%\email{t.shimaya@noneq.phys.s.u-tokyo.ac.jp}
\affiliation{Department of Physics,\! The University of Tokyo,\! 7-3-1 Hongo,\! Bunkyo-ku,\! Tokyo 113-0033,\! Japan}%
\author{Kazumasa A. Takeuchi}
%\email{kat@kaztake.org}
\affiliation{Department of Physics,\! The University of Tokyo,\! 7-3-1 Hongo,\! Bunkyo-ku,\! Tokyo 113-0033,\! Japan}%
\affiliation{Institute for Physics of Intelligence,\! The University of Tokyo,\! 7-3-1 Hongo,\! Bunkyo-ku,\! Tokyo 113-0033,\! Japan}%

\maketitle

\section{Supplementary figures}

\begin{figure}[h]
    \centering
    \includegraphics[width=0.85\textwidth]{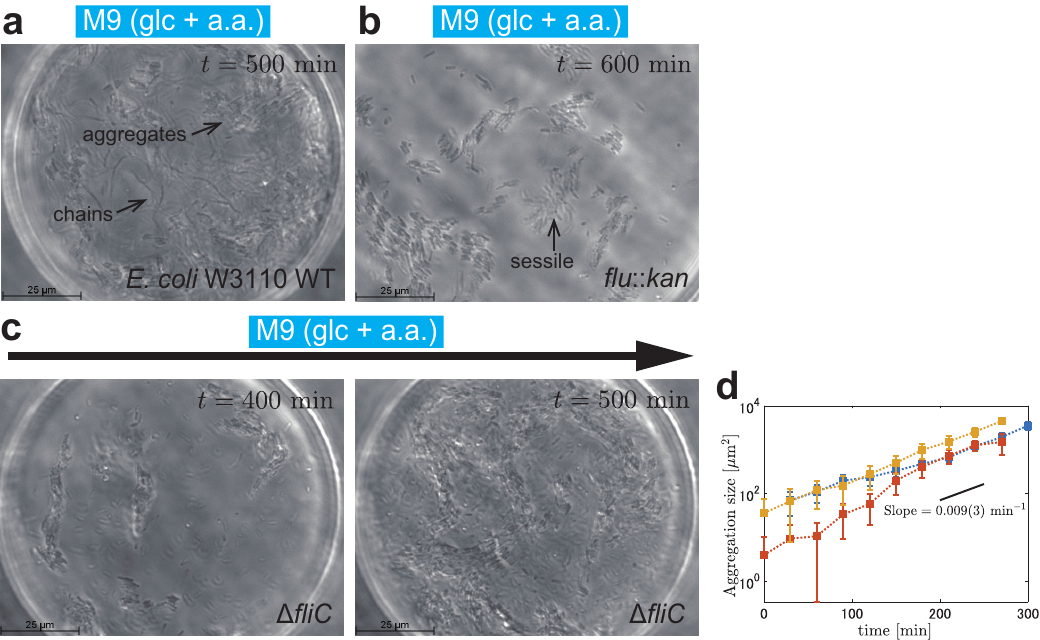}
    \caption{Aggregation processes of planktonic \textit{E. coli} cells in steady growth environments.
    (a) Populations of the wild-type W3110 (WT) in a well (diameter $\phi=110\unit{\mu{}m}$) cultured with M9(glc+a.a.) for $500\unit{min}$ since the observation was started.
    Initially there were 5 cells in the well.
    The appearance of chains and aggregates was confirmed in all 30 wells of various diameters recorded in a single experiment (WT growth in \tbref{M-tb-2}).
    See also \vidref1.
    (b) Aggregates of the \textit{flu}::\textit{kan} mutant (without Ag43 expression), in a well ($\phi=150\unit{\mu{}m}$) cultured with M9(glc+a.a.) for 600 min since the observation was started.
    Initially there were 10 isolated cells in the well.
    No chain formation was observed in 23 out of 28 wells with various diameters recorded in a single experiment (\textit{flu}::\textit{kan} growth in \tbref{M-tb-2}).
    In the other 5 wells, chain formation was observed, which may be due to flagella as mentioned below.
    In this image, cells adhering to the substrate (slightly out of the focus) can be seen.
    See also \vidref2.
    (c) An aggregation process of the $\Delta$\textit{fliC} mutant (without flagella) in a well ($\phi=110\unit{\mu{}m}$) cultured with M9(glc+a.a.).
    $t=0$ is the time at which the observation was started.
    Initially there were 8 cells in the well.
    No chain formation was observed in 28 out of 30 wells with various diameters recorded in a single experiment ($\Delta$\textit{fliC} growth in \tbref{M-tb-2}).
    In the other 2 wells, chain formation was observed.
    This suggests that both flagella and Ag43 are important for chain formation, though they can sometimes complement each other.
    While the adhesion of cells to the substrate was found in 27 out of 30 wells for the WT and in 23 out of 28 wells for the \textit{flu}::\textit{kan} mutant, this happened only in 8 out of 30 wells for the $\Delta$\textit{fliC} mutant.
    See also \vidref3.
    (d) Time evolution of the aggregate size of the $\Delta$\textit{fliC} mutant.
    Each time series represents the data in a separate well ($\phi=110\unit{\mu{}m}$).
    The error was estimated to be the perimeter $\times$ 5 pixels ($\sim 0.85\unit{\mu{}m}$), which we considered to be the range of uncertainty in manual detection of the edge of the regions.
    From the slope of the semi-log plot, we evaluated the growth rate at $0.009(3)\unit{min}^{-1}$.
    }
    \label{figS1}
\end{figure}

\begin{figure}[h]
    \centering
    \includegraphics[width=\textwidth]{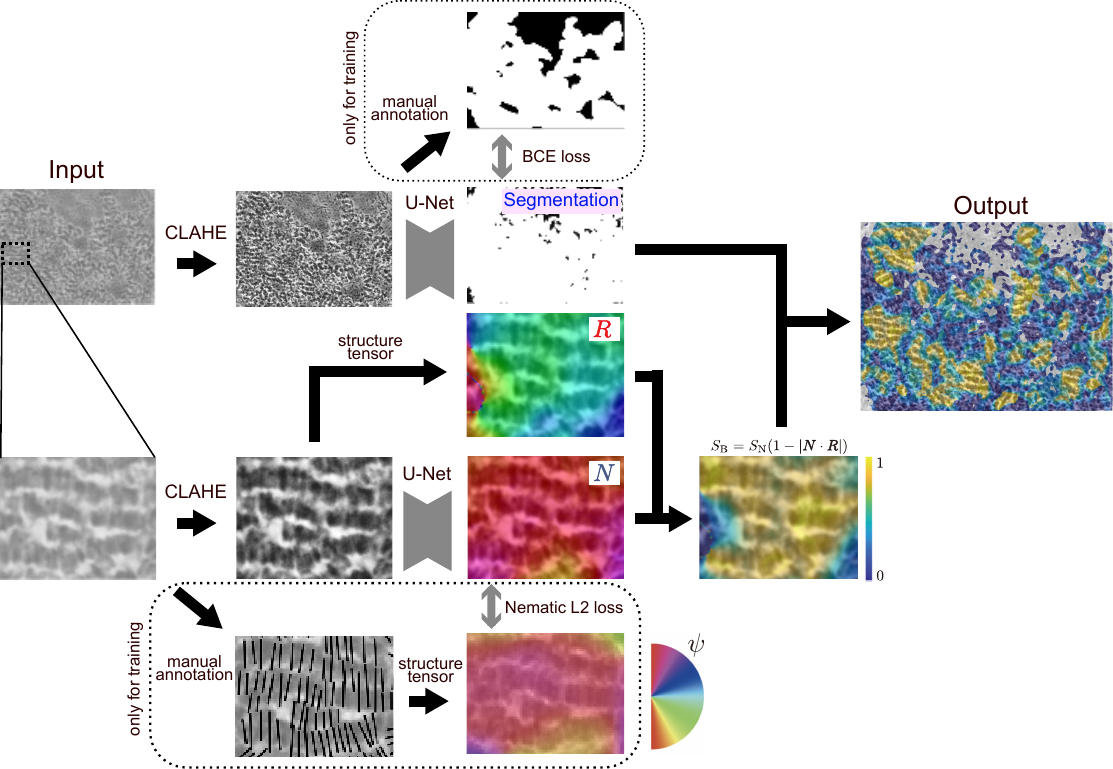}
    \caption{The entire pipeline of the image analysis.
    A deep neural network, namely the U-Net, was employed in two places, specifically for the segmentation of the regions with planktonic aggregates and for the estimation of the cell orientation $\bm{N}$.
    Note that entire images were given as the input of the U-Net for estimating $\bm{N}$ while the figure shows a small part of the images for visibility.
    See also Materials and Methods in the main text for details.
    }
    \label{figS2}
\end{figure}

\begin{figure}[h]
    \centering
    \includegraphics[width=\textwidth]{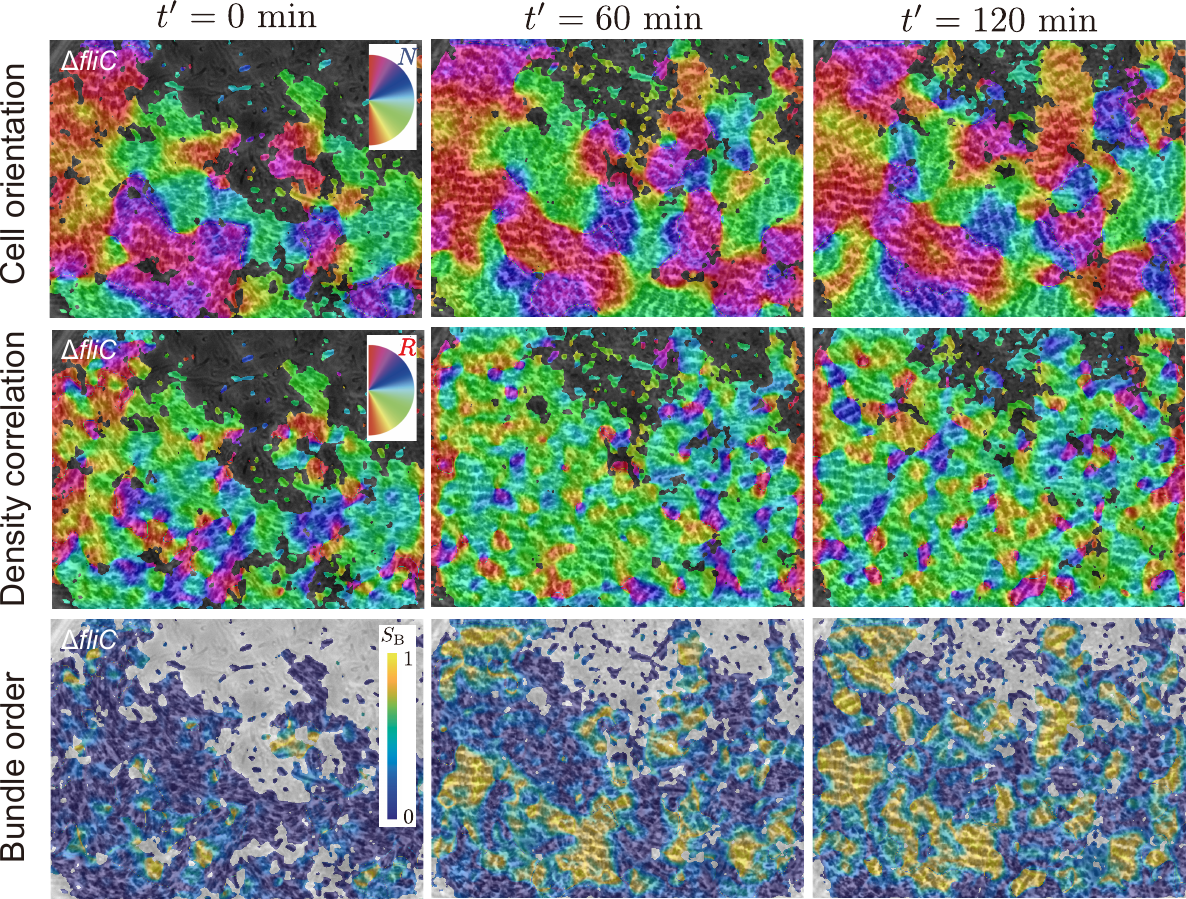}
    \caption{Cell orientation $\bm{N}$, density correlation director $\bm{R}$, and the bundle order parameter $S_\mathrm{B}$, overlaid on the microscope images.
    These are the results for the $\Delta$\textit{fliC} mutant grown in M9(glc+a.a.) and starved in PBS ($\Delta$\textit{fliC} starvation in \tbref{M-tb-2}).
    $\bm{N}$ and $\bm{R}$ are perpendicular if bundles are formed.
    The spatially-averaged bundle order parameter is shown in \figref{M-fig-3}d by the purple curve.
    See also \vidsref{5 and 6}.
    }
    \label{figS3}
\end{figure}

\begin{figure}[h]
    \centering
    \includegraphics[width=0.75\textwidth]{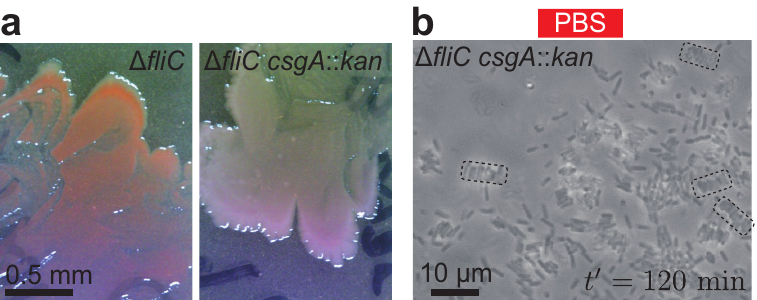}
    \caption{
    Investigation of the relevance of curli to the bundle formation.
    (a) Confirmation of the loss of curli in the $\Delta$\textit{fliC} \textit{csgA}::\textit{kan} mutant, by congo red.
    This dye selectively stains amyloid fibers including curli, as confirmed for the $\Delta$\textit{fliC} strain (left).
    No signal was detected for the $\Delta$\textit{fliC} \textit{csgA}::\textit{kan} strain (right).
    (b) Starvation response of the $\Delta$\textit{fliC} \textit{csgA}::\textit{kan} mutant grown in M9(glc+a.a.) and starved in PBS.
    Bundle clusters are indicated by the dashed rectangles.
    The well diameter $\phi$ is $230\unit{\mu{}m}$.
    $t'=0$ is the time at which the environment was switched.
    The bundle formation was observed in 15 out of 19 wells with various diameters recorded in a single experiment ($\Delta$\textit{fliC} \textit{csgA}::\textit{kan} starvation in \tbref{M-tb-2}).
    See also \vidref{11}.
    }
    \label{figS4}
\end{figure}

\begin{figure}[h]
    \centering
    \includegraphics[width=\textwidth]{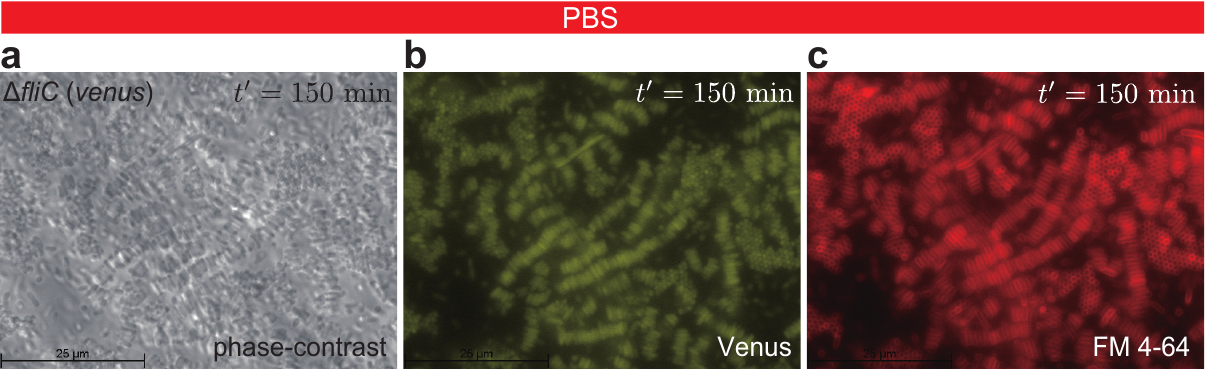}
    \caption{
    Multi-channel observation of the bundle formation of the $\Delta$\textit{fliC} (\textit{venus}) mutant, $150\unit{min}$ after the environment was switched from M9(glc+a.a.) to PBS.
    (a) Phase-contrast image. (b) Venus fluorescence image, which visualizes the entire cell bodies. (c) FM 4-64 fluorescence image, which visualizes the cell membranes and OMVs.
    %No localization of OMVs between bundle clusters was observed.
    }
    \label{figS5}
\end{figure}

\begin{figure}[h]
    \centering
    \includegraphics[width=\textwidth]{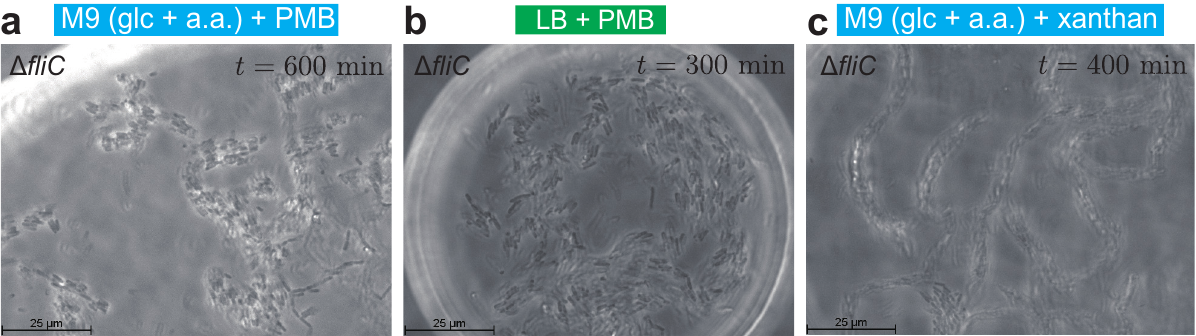}
    \caption{
    Steady growth measurements intended to test the relevance of depletion attraction.
    $t=0$ is the time at which the observation was started.
    (a) Observation of the $\Delta$\textit{fliC} mutant grown in M9(glc+a.a.) with $250\unit{ng/ml}$ polymyxin B.
    The well diameter $\phi$ is $230\unit{\mu{}m}$.
    Bundle clusters were not observed in all 6 wells with various diameters recorded in a single experiment ($\Delta$\textit{fliC} starvation (PMB) in \tbref{M-tb-2}).
    See also \vidref{12}.
    (b) Observation of the $\Delta$\textit{fliC} mutant grown in LB with $250\unit{ng/ml}$ polymyxin B.
    The well diameter $\phi$ is $110\unit{\mu{}m}$.
    Bundle clusters were not observed in all 15 wells with various diameters recorded in a single experiment ($\Delta$\textit{fliC} starvation (LB, PMB) in \tbref{M-tb-2}).
    See also \vidref{13}.
    (c) Observation of the $\Delta$\textit{fliC} mutant grown in M9(glc+a.a.) with $0.03\unit{wt\%}$ xanthan gum.
    The well diameter $\phi$ is $150\unit{\mu{}m}$.
    Bundle clusters were not observed in all 6 wells with various diameters recorded in a single experiment ($\Delta$\textit{fliC} starvation (xanthan) in \tbref{M-tb-2}).
    See also \vidref{14}.
    }
    \label{figS6}
\end{figure}

%\section{Supplementary tables}

\clearpage
\section{Movie descriptions}
\noindent
{\bf \vidref1:}\\
Growth of the wild-type W3110 (WT) in M9(glc.+a.a.).
The diameter of the well is $110\unit{\mu{}m}$.
The movie is played 1200 times faster than the real speed.\\
{\bf \vidref2:}\\
Growth of the \textit{flu}::\textit{kan} mutant in M9(glc.+a.a.).
The diameter of the well is $150\unit{\mu{}m}$.
The movie is played 1200 times faster than the real speed.\\
{\bf \vidref3:}\\
Growth of the $\Delta$\textit{fliC} mutant in M9(glc.+a.a.).
The diameter of the well is $110\unit{\mu{}m}$.
The movie is played 1200 times faster than the real speed.\\
{\bf \vidref4:}\\
Starvation response of the $\Delta$\textit{fliC} mutant.
The environment was switched from M9(glc.+a.a.) to PBS at $t=480\unit{min}$.
The diameter of the well is $150\unit{\mu{}m}$.
The movie is played 1200 times faster than the real speed.\\
{\bf \vidref5:}\\
The density correlation director $\bm{R}$ (top) and the cell orientation $\bm{N}$ (bottom) overlaid on \vidref{4}.
The pseudo-color indicates the angle (see \figref{M-fig-3}b, \figref{figS2} or \figref{figS3} for the definition).
$\bm{N}$ and $\bm{R}$ are perpendicular in the regions where smectic-like bundles are formed.\\
{\bf \vidref6:}\\
The bundle order parameter $S_\mathrm{B}$ overlaid on \vidref{4}.
The pseudo-color indicates the value of the bundle order parameter.
See \figref{M-fig-3}c for the color bar.\\
{\bf \vidref7:}\\
Starvation response of the $\Delta$\textit{fliC} mutant.
The environment was switched from LB to PBS at $t=225\unit{min}$.
The diameter of the well is $110\unit{\mu{}m}$.
The movie is played 1200 times faster than the real speed.\\
{\bf \vidref8:}\\
Starvation response of the wild-type W3110 (WT).
The environment was switched from M9(glc.+a.a.) to PBS at $t=480\unit{min}$.
The diameter of the well is $150\unit{\mu{}m}$.
The movie is played 1200 times faster than the real speed.\\
{\bf \vidref9:}\\
Starvation response of the $\Delta$\textit{fliC} mutant, grown in M9(glc.+a.a.) + $2\unit{\mu{}M}$ BisBAL and starved in PBS + $2\unit{\mu{}M}$ BisBAL.
Bundles did not appear in this movie.
The environment was switched at $t=480\unit{min}$.
The diameter of the well is $150\unit{\mu{}m}$.
The movie is played 1200 times faster than the real speed.\\
{\bf \vidref{10}:}\\
Starvation response of the $\Delta$\textit{fliC} mutant), grown in M9(glc.+a.a.) + $2\unit{\mu{}M}$ BisBAL and starved in PBS + $2\unit{\mu{}M}$ BisBAL.
Bundles appeared in this movie.
The environment was switched at $t=480\unit{min}$.
The diameter of the well is $150\unit{\mu{}m}$.
The movie is played 1200 times faster than the real speed.\\
{\bf \vidref{11}:}\\
Starvation response of the $\Delta$\textit{fliC} \textit{csgA}::\textit{kan} mutant.
The environment was switched from M9(glc.+a.a.) to PBS at $t=480\unit{min}$.
The diameter of the well is $230\unit{\mu{}m}$.
The movie is played 1200 times faster than the real speed.\\
{\bf \vidref{12}:}\\
Growth of the $\Delta$\textit{fliC} mutant in M9(glc+a.a.) + $250\unit{ng/ml}$ polymyxin B.
The diameter of the well is $230\unit{\mu{}m}$.
The movie is played 1200 times faster than the real speed.\\
{\bf \vidref{13}:}\\
Growth of the $\Delta$\textit{fliC} mutant in LB + $250\unit{ng/ml}$ polymyxin B.
The diameter of the well is $230\unit{\mu{}m}$.
The movie is played 1200 times faster than the real speed.\\
{\bf \vidref{14}:}\\
Growth of the $\Delta$\textit{fliC} mutant in M9(Glc+a.a.) + $0.03\unit{wt\%}$ xanthan gum.
The diameter of the well is $150\unit{\mu{}m}$.
The movie is played 1200 times faster than the real speed.\\
{\bf \vidref{15}:}\\
Response against starvation and growth recovery of the $\Delta$\textit{fliC} mutant.
The environment was switched from M9(glc.+a.a.) to PBS at $t=480\unit{min}$, and from PBS to M9(glc.+a.a.) at $t=660\unit{min}$.
The diameter of the well is $230\unit{\mu{}m}$.
The movie is played 1200 times faster than the real speed.

%\bibliography{BDpaper}